\documentclass[preprint]{aastex}
\usepackage{amsbsy} 

\newcommand{\Msun}{M_\odot}

\newcommand{\bx}{\boldsymbol{x}}
\newcommand{\by}{\boldsymbol{y}}

\slugcomment{submitted to MNRAS}

\shorttitle{The Approach to Collapse of Molecular Clouds}
\shortauthors{Stahler \& Yen}

\begin{document}

\title{The Approach to Collapse of Molecular Clouds}

\author{Steven W. Stahler\altaffilmark{1} and Jeffrey J. Yen\altaffilmark{2}}

\altaffiltext{1}{Astronomy Department. University of California,
Berkeley, CA 94720}
\altaffiltext{2}{Physics Department, University of California,
Berkeley, CA 94720}

\email{Sstahler@astro.berkeley.edu}

\begin{abstract}
The dense molecular cloud cores that form stars, like other self-gravitating 
objects, undergo bulk oscillations. Just at the point of gravitational 
instability, their fundamental oscillation mode has zero frequency. We study, 
using perturbation theory, the evolution of a spherical cloud that possesses 
such a frozen mode. We find that the cloud undergoes a prolonged epoch of 
subsonic, accelerating contraction. This slow contraction occurs whether 
the cloud is initially inflated or compressed by the oscillation. The subsonic 
motion described here could underlie the spectral infall signature observed in 
many starless dense cores.  
\end{abstract}


\section{Introduction}

The central process in star formation is the gravitational collapse of a dense,
molecular cloud core. Such objects are well studied empirically, through a
variety of techniques. Less than half of observed dense cores already contain
embedded stars, with the fraction increasing at higher mean column density and
volume density \citep{b86,jw00}. The remaining, so-called starless, cores 
exhibit a range of density contrasts and temperature gradients 
\citep[e.g.,][]{t04,c07}. Those of higher density do not differ in their gross 
properties from cores with stars \citep{lm99}. Observations, then, have not 
yet provided a clearcut answer to the salient question: How does the collapse 
process actually begin?

One important clue is the low level of internal motion observed within dense 
cores. This fact was established early, from the relatively narrow linewidths
of optically thin molecular tracers such as NH$_3$ \citep{mb83}. The linewidths
reveal gas motion that is largely subsonic, in contrast to the turbulent or 
wave-like character of the cores' external environments \citep{g98}. Other 
studies bolster the view from spectroscopy. \citet{a01} used near-infrared 
extinction mapping to reconstruct the density structure of the starless core 
B68. After azimuthal averaging, the density profile is close to a theoretical 
one for an isothermal sphere in balance between self-gravity and gas pressure 
\citep{e55,b56}. Thus, the object appears to be in dynamical equilibrium, as 
expected when internal motion is subsonic. Similar results have since been 
found for a number of other starless cores \citep{t04,k05}. The inferred 
density profiles are consistent with earlier ones reconstructed through
submillimeter emission maps \citep{wa99}.

Real dense cores are not perfect spheres, in part because they derive 
additional support from the anisotropic force associated with the interstellar
magnetic field \citep{c99}. In any case, the fact that clouds begin collapse
from a near-equilibrium state is significant theoretically. What is the nature
of this equilibrium? Clearly, the initial state cannot be dynamically 
{\it stable} with respect to small perturbations, as no collapse would 
ensue.\footnote{Starless cores of the lowest density contrast are indeed 
stable, and are confined by external pressure more than self-gravity. 
\citet{a08} and \citet{kc08} have suggested that many such objects are fated 
never to collapse, but will gradually disperse.}  On the other hand, it is 
also unlikely to be dynamically {\it unstable}; it is difficult to imagine the 
prehistory of an object that could have veered away sharply from equilibrium at
any time. The most natural initial state just prior to collapse for a 
star-forming dense core is one of {\it marginal stability}. While the object 
is in force balance, its fundamental mode of oscillation has zero frequency. 
Working again within the idealized framework of isothermal spheres, our paper 
addresses two specific questions. How does a marginally stable, isothermal 
cloud evolve with time? Can we account in a simple and compelling way for the 
tendency of such an object to collapse?

Collapse calculations have a long history, and we are by no means the first to
recognize the particular importance of the marginally stable isothermal sphere 
as an initial state. \citet{h77}, \citet{fc93}, \citet{o99}, and \citet{a05} 
all followed the collapse of such a configuration. However, these authors were 
concerned with developments at a relatively advanced stage. The subtle question
of how the cloud first evolves away from equilibrium was effectively bypassed 
through assuming that the cloud was initially slightly overdense with respect 
to equilibrium, thereby guaranteeing eventual collapse. Our calculation 
focuses entirely on the transition issue. We show, using the tools of 
perturbation theory, that marginally stable clouds enter a protracted phase 
of slow, but accelerating inward contraction. We do not follow the evolution 
deep into collapse, a task that could be pursued using more standard methods.

Our study of this early contraction phase holds more than purely theoretical
interest. A large fraction of starless cores show convincing signs of inward
motion \citep{wi99,lmt99,ge00,lmt01,s07}. The spectroscopic signature of 
infall is an asymmetric emission line, often self-absorbed, that is skewed 
toward the blue. Inferred speeds are well below thermal, but greater than 
those associated with gravitational settling mediated by ambipolar diffusion 
of the magnetic field \citep[as calculated, e.g., by][]{cm94}. Our accelerating
contraction is an attractive candidate to explain this intriguing, and perhaps 
pivotal, set of observations.

In Section~2 below, we present our method of solution. After 
nondimensionalization of the dynamical equations, we introduce a perturbation
expansion that allows us to separate out the first-order, oscillatory motion
from the second-order displacement that progressively evolves with time. 
Sections~3 and 4 are devoted to both analytical derivations and numerical 
results. In the first, we display the cloud's normal modes of oscillation, 
including the critical one of zero frequency. In the second, we trace the bulk 
contraction of a cloud subject to the zero-frequency mode. Finally, Section~5 
discusses the possible connection to the infall signature of starless cores, 
and indicates fruitful extensions of this work.

\section{Solution Strategy}
\subsection{Physical Assumptions}

Our task of following early cloud evolution is facilitated by adopting a
simplified physical picture. These simplifications are by now traditional, but
it is important to revisit them as new observations arise and theoretical
understanding grows. Thus, our employment of an isothermal equation of state,
with a gas temperature that is constant in both space and time, is technically
inconsistent with recent observational studies of starless cores. The inferred
temperature of L1544, a starless core in Taurus that exhibits infall, decreases
from 12~K at the outer edge to 6~K near the center \citep{c07}. The central
temperature is depressed by the partial shielding of dust grains from external 
starlight; these grains are thermally coupled to the gas through collisions. 
However, the region that is effectively shielded comprises little of the total 
mass, and collapse calculations that account for the finite temperature 
gradient show it to have a relatively minor effect \citep{kf05}.
      
Another traditional assumption, and one ostensibly even more radical, is that 
our model cloud is spherical. It has long been known that the projected 
molecular-line emission contours of starless cores, like dense cores in 
general, are more accurately elliptical, with mean axis ratios of about 2:1
\citep{lm99}. The deprojected, or intrinsic, distribution of shapes must be
inferred statistically from observations of a large sample. Under the 
condition that the underlying structure be axisymmetric, a prolate 
configuration best matches the data \citep{m91,r96}. Relaxing this restriction,
\citet{g02} found that a triaxial configuration is preferred
\citep[see also][]{j01}. This result is puzzling. As we stated earlier, a
departure from spherical symmetry could partially be explained by additional
magnetic support. However, \citet{g05} has demonstrated that there are no
triaxial magnetostatic structures, at least for scale-free configurations.
It may be, as Galli suggests, that the observed shapes reflect distortion 
created by internal oscillations, a phenomenon we ourselves shall invoke 
shortly. In any event, our adoption of a spherical geometry, along with our 
neglect of magnetic support, can only be viewed as a convenient, first 
approximation. The essential finding of accelerating contraction from marginal 
stability should continue to hold within a more complete description of the 
equilibrium, magnetostatic state.

Within the last decade, a number of researchers have questioned the very 
existence of equilibrium structures. They have made the assertion, based on
numerical simulations, that dense cores are transient, fully dynamic entities.
Such studies, recently reviewed by \citet{b07}, treat an interior portion of 
the larger, parent cloud as a computational box filled with gas, with or 
without an embedded magnetic field. Turbulence is simulated by stirring the 
gas. This energy input is eventually offset by the unavoidable numerical 
dissipation present in the calculation. Once such a steady state is attained,
external stirring is turned off and self-gravity turned on. Sufficiently
overdense structures collapse on themselves, in apparent imitation of the
star formation process.\footnote{In some simulations, turbulence is 
continually driven, even after self-gravity is turned on \citep[e.g.,][]{v05}.}

It is intriguing that these structures resemble, in their masses, sizes, and
geometric aspect ratios, observed dense cores. Even triaxial structures have
been obtained \citep[e.g.,][]{o08}. However, the key observational distinction
between dense cores and their surroundings is their subsonic (or, more
properly, sub-Alfv\'enic) internal velocity. This fact implies, as we have
stressed, that the entities, unlike their numerical surrogates, are in force
balance. Moreover, the large observed fraction of starless cores indicates a
correspondingly protracted epoch in a typical dense core's lifetime prior to
collapse \citep{lm99,jw00,ki05}. If these structures indeed condense from a 
turbulent medium, their initial growth may be retarded by a relatively strong 
ambient magnetic field \citep{g07}. 

\subsection{Dynamical Equations and their Nondimensionalization}

We consider a spherical cloud if fixed mass $M$ and isothermal sound speed
$a$. The equation of state relating the internal pressure $P$ and density
$\rho$ is
\begin{equation}
P\,=\,\rho\,a^2 \,\,.
\end{equation}
The outer edge of the cloud is set by the condition that $P$ falls to some 
external value $P_e$, also constant. In spherical symmetry, any matter 
outside the boundary exerts no gravitational force on the cloud. Hence we do 
not need to consider this medium in detail, beyond the fact that it exerts a 
fixed pressure. In particularly, we need not assume (unrealistically) that the
external gas has a relatively high temperature and low density, as is 
frequently done in collapse calculations \citep[e.g.,][]{fc93}. We shall see
that the cloud boundary, as just defined, shrinks during the evolution. This
finding suggests that the distinction between ``cloud'' and ``external medium''
is somewhat artificial, and that inward motion is spatially extended. Indeed,
the observed infall signature of starless cores is striking for its broad
extent \citep{m00}.

In Eulerian coordinates $(r,t)$, the equation of mass continuity is
\begin{equation}
{{\partial\rho}\over{\partial t}} \,=\, -{1\over r^2}\,
{{\partial(r^2\,\rho\,u)}\over{\partial r}} \,\,,
\end{equation} 
where $u (r,t)$ is the fluid velocity. This velocity satisfies momentum
conservation:
\begin{equation}
{{\partial u}\over{\partial t}} \,+\, 
u\,{{\partial u}\over{\partial r}} \,=\,
-{a^2\over\rho}\,
{{\partial\rho}\over{\partial r}} \,-\,
{{G\,m}\over{r^2}} \,\,.
\end{equation}
Here, $m$ is the mass enclosed within any radius:
\begin{equation}
m \,\equiv\, \int_0^r\!4\,\pi\,r^2\,\rho\,dr \,\,.
\end{equation}

As the total cloud mass is fixed, we shall find it more convenient to switch to
Lagrangian coordinates $(m,t)$. The mass continuity equation~(2) becomes 
\begin{equation}
{{\partial\rho}\over{\partial t}} \,=\, -4\,\pi\,\rho^2 \,
  {{\partial{\phantom u}}\over{\partial m}} \left( r^2\,
{{\partial r}\over{\partial t}}\right) \,\,,
\end{equation}
Here, we have replaced the velocity $u$ by the time-derivative of the (now
dependent) variable $r$:
\begin{equation}
u \,=\, {{\partial r}\over{\partial t}} \,\,,
\end{equation}
and have again suppressed all subscripts in the partial derivatives. The 
momentum equation~(3) transforms to
\begin{equation}
{{\partial^2 r}\over{\partial t^2}} \,=\, -4\,\pi\,r^2\,a^2\,
{{\partial\rho}\over{\partial m}} \,-\,
{{G\,m}\over r^2}\,\,.
\end{equation}
Finally, equation~(4) for $m$ is replaced by
\begin{equation}
{{\partial r}\over{\partial m}} \,=\,
{1\over{4\,\pi\,r^2\,\rho}}\,\,.
\end{equation}

Equations (5), (7), and (8) constitute our basic dynamical equations. We may 
further cast them into nondimensional form using the three constants $G$, $a$, 
and $P_e$. We define nondimensional versions of $m$, $r$, $\rho$, and $t$:
\begin{eqnarray}
{\tilde m} &\,\equiv\,& {{G\,m\,\sqrt{4\,\pi\,G\,P_e}}\over a^4} \,\,, \\
{\tilde r} &\,\equiv\,& {{r\,\sqrt{4\,\pi\,G\,P_e}}\over a^2} \,\,, \\
{\tilde\rho} &\,\equiv\,& {{\rho\,a^2}\over P_e} \,\,, \\
{\tilde t} &\,\equiv\,& {{t\,\sqrt{4\,\pi\,G\,P_e}}\over a} \,\,. 
\end{eqnarray}
We further replace $\tilde\rho$ by a new nondimensional variable $\tilde\phi$,
defined through
\begin{equation}
{\tilde\rho} \,\equiv\,{\rm e}^{-{\tilde\phi}} \,\,. 
\end{equation}
After dropping the tilde, our dynamical equations become
\begin{eqnarray}
{{\partial\phi}\over{\partial t}} &\,=\,& {\rm e}^{-\phi}\,
{{\partial{\phantom u}}\over {\partial m}}
\left(r^2\,{{\partial r}\over{\partial t}}\right) \,\,, \\
{{\partial^2 r}\over{\partial t^2}} &\,=\,& r^2\,{\rm e}^{-\phi} 
{{\partial\phi}\over{\partial m}} \,-\, {m\over r^2} \,\,, \\
{{\partial r}\over{\partial m}} &\,=\,& {{{\rm e}^\phi}\over r^2} 
\,\,.
\end{eqnarray}
We will also be using equation~(6) relating $u$ and $r$. If we set
\hbox{${\tilde u}\,\equiv\,u/a$}, then this equation remains the same   
nondimensionally. 

\subsection{Perturbation Expansion}

Equations (14)-(16) are to be solved subject to the inner boundary condition
\hbox{$r(0,t)\,=\,0$} and the outer one \hbox{$\phi(M,t)\,=\,0$}. The latter
is just the requirement, expressed in nondimensional language, that 
the gas pressure at the edge equal the fixed value $P_e$. We must also specify
the cloud's initial configuration. This state is itself slightly perturbed
from true equilibrium, under the influence of the oscillation mode. We are thus
motivated to introduce a perturbation expansion of the dependent variables 
about the equilibrium state. We write:  
\begin{eqnarray}
\phi(m,t) &\,=\,& \phi_0 (m) \,+\, \epsilon\,\phi_1 (m,t) \,+\,
\epsilon^2\,\phi_2 (m,t) \,+\, ... \,\,, \\
r(m,t) &\,=\,& r_0 (m) \,+\, \epsilon\,r_1 (m,t) \,+\,
\epsilon^2\,r_2 (m,t) \,+\, ... \,\,.
\end{eqnarray}
Here, $\epsilon$ is a small, free parameter. The physical significance of
$\epsilon$ is that it will set the amplitude of the oscillation mode present 
at the outset (see \S 4 below). Quantities with the 0 subscript refer to the 
time-independent, equilibrium state. 

We are now in a position to separate out the hierarchy of cloud motions. We
substitute our series expansions into equations (14) - (16) and equate like 
powers of $\epsilon$. At the lowest (zeroth) order in $\epsilon$, both sides 
in the equation of mass continuity vanish. However, equations (15) and (16) 
yield, to the same order, the structural equations of the underlying, 
equilibrium state:
\begin{eqnarray}
{{d\phi_0}\over{dm}} &\,=\,& {{m\,{\rm e}^{\phi_\circ}}\over r_0^4}  
\,\,,\\
{{dr_0}\over{dm}} &\,=\,& {{{\rm e}^{\phi_\circ}}\over r_0^2} \,\,.
\end{eqnarray}

When we equate terms proportional to $\epsilon$, all three equations give a 
non-trivial result:
\begin{eqnarray}
{{\partial\phi_1}\over{\partial t}} &\,=\,&
{\rm e}^{-\phi_\circ}\,
{{\partial{\phantom u}}\over{\partial m}}  
\left(r_0^2\,{{\partial r_1}\over{\partial t}}\right)
\,\,,\\
{{\partial^2 r_1}\over{\partial t^2}} &\,=\,& {\rm e}^{-\phi_\circ} 
\,{{d\phi_0}\over{dm}}
\left(2\,r_0\,r_1 \,-\, r_0^2\,\phi_1 \right) \,+\,
r_0^2\,{\rm e}^{-\phi_\circ}
\,{{\partial\phi_1}\over{\partial m}} \,+\,
{{2\,m\,r_1}\over r_0^3}
\,\,,\\
{{\partial r_1}\over{\partial m}} &\,=\,& 
{{\rm e}^{\phi_\circ}\over r_0^2} 
\left(\phi_1 \,-\, {{2\,r_1}\over r_0}\right) \,\,.
\end{eqnarray} 

These zeroth- and first-order equations are sufficient to describe small 
oscillations about a general equilibrium state. However, we are interested in 
that unique equilibrium for which the fundamental oscillation frequency 
vanishes. The perturbed cloud will not be perfectly static, but will evolve
slowly, through second-order effects. Thus, we also need to equate terms 
in the dynamical equations proportional to $\epsilon^2$. We find
\begin{eqnarray}
\phi_1\,{{\partial\phi_1}\over{\partial t}} \,+\, 
{{\partial\phi_2}\over{\partial t}} &\,=\,& {\rm e}^{-\phi_\circ}\,
{{\partial{\phantom u}}\over{\partial m}} 
\left(2\,r_0\,r_1\,{{\partial r_1}\over{\partial t}} \,+\,
r_0^2\,{{\partial r_2}\over{\partial t}}
\right)\,\,,\\
{{\partial^2 r_2}\over{\partial t^2}} &\,=\,& {\cal A}_1 \,+\, {\cal A}_2
\,\,, \\
{{\partial r_2}\over{\partial m}} &\,=\,& 
{{{\rm e}^{\phi_\circ}}\over r_0^2}\,
\left({{3\,r_1^2}\over r_0^2} \,-\, 
{{2\,r_2}\over r_0} \,+\, \phi_2 \,-\,
{{2\,\phi_1\,r_1}\over r_0} \,+\, {\phi_1^2\over 2} \right) \,\,.
\end{eqnarray}   
The two acceleration terms ${\cal A}_1$ and ${\cal A}_2$ in equation~(25) are
\begin{eqnarray}
{\cal A}_1 &\,=\,& {\rm e}^{-\phi_\circ}\,{{d\phi_0}\over{dm}}
\left( {{r_0^2\,\phi_1^2}\over 2} \,-\, r_0^2\,\phi_2 \,+\, r_1^2 \,-\,
2\,r_0\,r_1\,\phi_1 \,+\, 2\,r_0\,r_2 \right) \,\,, \\
{\cal A}_2 &\,=\,& {\rm e}^{-\phi_\circ}\, 
{{\partial\phi_1}\over {\partial m}}\,
\left( 2\,r_0\,r_1 \,-\, r_0^2\,\phi_1 \right) \,+\,
r_0^2\,{\rm e}^{-\phi_\circ}\,{{\partial\phi_2}\over{\partial m}} \,+\,
{{2\,m\,r_2}\over r_0^3} \,-\,{{3\,m\,r_1^2} \over r_0^4} \,\,.
\end{eqnarray} 

\section{Cloud Oscillations}
\subsection{Equilibria}

We consider first equations (19) and (20), that describe the equilibrium state.
These equations are subject to the boundary conditions 
\hbox{$r_0 (0) \,=\, 0$} and \hbox{$\phi_0 (M) \,=\, 0$}, to be 
applied at the center and outer edge, respectively. More accurately, the 
second condition {\it defines} the boundary of the equilibrium configuration. 
As is well known, there exists a one-parameter family of such structures, each 
characterized by its own center-to-edge density contrast. We denote this 
contrast as $\rho_c$, since it is identical to the nondimensional central 
density in our formulation.

Because of the inner boundary condition, the righthand sides of both equations
are singular at the origin. To start our numerical integration, we therefore 
expand both $r_0 (m)$ and $\phi_0 (m)$ in the appropriate power series:
\begin{eqnarray}
r_0 (m) &\,=\,&  a_0\,m^{1/3} \,+\, a_1\,m \,+\, a_2\,m^{5/3} 
\, + \, ... \,\,, \\
\phi_0 (m) &\,=\,& -{\rm ln}\,\rho_c \,+\, b_0\,m^{2/3} \,+\, 
b_1\,m^{4/3}\,+\, b_2\,m^2  \,+\, ... \,\,.
\end{eqnarray}
The first two coefficients are
\begin{eqnarray}
a_0 &\,=\,& \left(3\over\rho_c\right)^{1/3} \,\,, \\
b_0 &\,=\,& {3\over{2\,\rho_c\,a_0^4}} \,\,,
\end{eqnarray}
and higher ones are determined recursively. 

Based on our numerical integration, the dashed curves in Figure~1 are density 
profiles for states with \hbox{$\rho_c\,=\,5$} and \hbox{$\rho_c \,=\, 30$}. 
Note that we have plotted 
\hbox{$\rho_0 \,\equiv\,{\rm exp}(-\phi_\circ)$} 
as a function of $r_0$, in the conventional manner. The middle, solid 
curve corresponds to \hbox{$\rho_c \,=\, 14.04$}, for which 
\hbox{$M\,=\,4.19$}. This is the famous Bonnor-Ebert state. As we shall verify 
shortly, it is the starting configuration of interest for the present problem. 
The other two curves represent outer limits to the empirical fitting of 
density profiles for most starless dense cores \citep{k05}.

\subsection{Normal Modes}

Equations (21)-(23) describe, to linear order, internal motion of the 
equilibrium cloud. All such motion can be decomposed into a series of normal
modes. The fundamental, or breathing, mode is generally expected to have the
largest amplitude, but higher harmonics may also be present 
\citep[as well as non-radial oscillations; see][]{k06}.

To obtain the full set of spherical normal modes, we first solve equation~(23) 
for the density perturbation:
\begin{equation}
\phi_1 \,=\, r_0^2\,
{\rm e}^{-\phi_\circ}\,
{{\partial r_1}\over{\partial m}} \,+\, 
{{2\,r_1}\over r_0} \,\,.
\end{equation}
We substitute this expression into the momentum equation~(22) and obtain a
partial differential equation for $r_1 (m,t)$:
\begin{equation}
{{\partial^2 r_1}\over {\partial t^2}} \,=\,
r_0^4\,{\rm e}^{-2\,\phi_\circ}\,
{{\partial^2 r_1}\over{\partial m^2}} \,+\,
2\,{\rm e}^{-\phi_\circ}\left( 2\,r_0 \,-\, m\right)
{{\partial r_1}\over {\partial m}} \,+\,
\left( {{2\,m} \over r_0^3} \,-\, {2\over r_0^2}\right) r_1 \,\,. 
\end{equation}
The coefficients in this equation are all functions of the chosen equilibrium
state.

Before proceeding with the solution of equation~(34), we note that the 
additional first-order equation~(21) has not been used in the derivation. In
fact, this relation is redundant, and already contained implicitly in
equation~(33). To see this, take the time derivative of the latter:
\begin{mathletters}
\begin{eqnarray} 
{{\partial\phi_1}\over{\partial t}} &\,=\,& r_0^2\,{\rm e}^{-\phi_\circ}\,
{{\partial^2 r_1}\over{{\partial t}\,{\partial m}}} \,+\,
{2\over r_0}\,{{\partial r_1}\over{\partial t}} 
\,\,,\\
{\phantom u} &\,=\,& r_0^2\,{\rm e}^{-\phi_\circ}\,
{{\partial^2 r_1}\over{{\partial t}\,{\partial m}}} \,+\,
2\,r_0\,{\rm e}^{-\phi_\circ}\,{{dr_0}\over{dm}}\,
{{\partial r_1}\over{\partial t}} \,\,, \\
{\phantom u} &\,=\,& {\rm e}^{-\phi_\circ}\,
{{\partial{\phantom u}}\over{\partial m}}
\left( r_0^2\,{{\partial r_1}\over{\partial t}}\right)
\,\,,
\end{eqnarray}
\end{mathletters}
which is just equation~(21). Here we have used the time-independence of all
equilibrium quantities, and have also employed equation~(20). 

Returning to equation~(34), we seek oscillatory solutions for $r_1 (m,t)$. 
Thus, we set
\begin{equation}
r_1 (m,t) \,=\, A\,\,\xi_1 (m)\,\,{\rm e}^{i\,\omega\,t} \,\,.
\end{equation}
Since equation~(34) is linear in $r_1$, the coefficient $A$ is an arbitrary 
constant. After substitution, we find an ordinary differential equation for 
$\xi (m)$:
\begin{equation}
0 \,=\, r_0^4\,{\rm e}^{-2\,\phi_0}\,{{d^2\xi_1}\over{dm^2}} \,+\,
2\,{\rm e}^{-\phi_\circ}\left( 2\,r_0\,-\,m\right)
{{d\xi_1}\over{dm}} \,+\,
\left( {{2\,m}\over r_0^3} \,-\, {2\over r_0^2} \,+\,\omega^2 \right)
\xi_1 \,\,.
\end{equation}
The boundary conditions are now \hbox{$\xi_1 (0) \,=\, 0$} and
\hbox{$\phi_1 (M) \,=\, 0$}. Using equation~(33), the latter may be transformed
into a condition on $\xi_1 (m)$:
\begin{equation}
{{d\xi_1}\over{dm}} \,=\,-{{2\,\xi_1\,{\rm e}^{\phi_\circ}}\over{r_0^3}}
\,\,.
\qquad{\it at}\,\,\, m\,=\,M
\end{equation}

We note that equation~(37) has a regular singular point at the origin. Once 
more, we start the integration through a power series development:
\begin{equation}
\xi_1 (m) \,=\, c_0\,m^{1/3} \,+\, c_1\,m \,+\, c_2\,m^{5/3} \,+\, ...
\,\,.
\end{equation}
The constant $c_0$ may be chosen arbitrarily. The next coefficient is
\begin{equation}
c_1 \,=\, -{{a_0^2\,c_0\,\omega^2}\over 10} \,\,,
\end{equation}
and successive ones may be similarly found.

For any central density $\rho_c$ of the equilibrium cloud, there exists a
sequence of $\omega^2$-values such that $\xi_1 (m)$ meets the two boundary
conditions. When \hbox{$\rho_c\,=\,14.04$}, the lowest value of $\omega^2$ is
zero. The corresponding $\xi_1 (m)$ is the oscillation mode of interest. If we
denote the fundamental, first, and second harmonics as $\xi_{10}$,
$\xi_{11}$, and $\xi_{12}$, then the corresponding $\omega^2$-values are
0, 8.37, and 24.3. Figure~2 plots these three normal modes, again using 
$r_0$ as the independent variable. In all cases, we have set the 
coefficient $c_0$ in equation~(39) to unity.
  
Note finally that we may use the equilibrium relations, equations~(19) and
(20), to rewrite equation~(37) as
\begin{equation}
0 \,=\, {{d{\phantom u}}\over{dm}}
\left( r_0^4\,{\rm e}^{-2\,\phi_\circ}\, {{d\xi_1}\over{dm}}\right) 
\,+\,
\left( {{2\,m} \over r_0^3} \,-\, {2\over r_0^2}\right)
\xi_1 \, +\,  \omega^2\,\xi_1 \,\,.
\end{equation}
This equation is of the Sturm-Liouville type. As expected physically, all of 
its eigenvalues $\omega^2$ are real. Distinct eigenmodes are orthogonal:
\begin{equation}
\int_0^M\!\xi_{1p}\,\xi_{1q}\,dm \,=\, 0 \,\,. 
\qquad{\it for}\,\,\, p\,\ne\,q
\end{equation} 
Since only the zero-frequency oscillation will be of interest in the following 
discussion, we will soon revert to the simpler notation $\xi_1 (m)$ for the 
fundamental mode, and assume the standard normalization \hbox{$c_0 \,=\, 1$} 
for this function. After also setting the coefficient $A$ in equation~(36) to 
unity, we will thus be making the identification 
\hbox{$r_1 (m,t) \,=\, \xi_1 (m)$}. 

\section{Cloud Contraction}
\subsection{Fundamental Equation}

Once we have selected the oscillatory mode of the equilibrium state, the
second-order equations (24)-(28) describe additional motion. This motion is,
of course, critical when the underlying oscillation has zero frequency, but we 
shall first keep the discussion more general. As in our derivation of the 
normal modes, it is convenient to solve for the density perturbation (now 
$\phi_2$) from the mass-radius relation, equation~(26): 
\begin{equation}
\phi_2 \,=\, r_0^2\,{\rm e}^{-\phi_\circ}\,{{\partial r_2}\over{\partial m}}
\,+\,{{2\,r_2}\over r_0} \,-\, {r_1^2\over r_0^2} \,-\,
{{r_0^4\,{\rm e}^{-2\,\phi_\circ}}\over 2}\,
\left({{\partial r_1}\over{\partial m}}\right)^2 \,\,, 
\end{equation}
where we have used equation~(33) for $\phi_1$. We substitute both this latter
relation and equation~(43) for $\phi_2$ into the righthand side of the 
momentum equation~(25). We thus derive our fundamental partial differential
equation for the displacement $r_2 (m,t)$:
\begin{equation}
{{\partial^2 r_2}\over{\partial t^2}} \,=\,
r_\circ^4\,{\rm e}^{-2\,\phi_\circ}\,{{\partial^2 r_2}\over{\partial m^2}} 
\,+\, 2\,{\rm e}^{-\phi_\circ} \left( 2\,r_0 \,-\, m\right)
{{\partial r_2}\over{\partial m}} \,+\,
\left( {{2\,m} \over r_0^3} \,-\, {2\over r_0^2}\right) r_2 \,+\, F \,\,,
\end{equation}
where
\begin{equation}
F \,\equiv\, \left( {2\over r_0^3}\,-\, {{3\,m}\over r_0^4}\right) r_1^2 \,+\,
{\rm e}^{-\phi_\circ}\left( {{2\,m}\over r_0}\,-\,2\right)
{{\partial r_1^2}\over{\partial m}} \,+\,
r_0^2\,{\rm e}^{-2\,\phi_\circ} \left( 2\,r_0\,-\,m\right)
\left( {{\partial r_1}\over{\partial m}}\right)^2 \,\,.
\end{equation}

We again remark that it has been unnecessary to invoke the continuity 
equation~(24). The explanation, as before, is that this equation yields no new
information. To see this, take the time derivative of equation~(43) and
multiply through by ${\rm e}^{\phi_\circ}$:
\begin{equation}
{\rm e}^{\phi_\circ}\,{{\partial \phi_2}\over{\partial t}}\,=\,
r_0^2\,{{\partial^2 r_2}\over{\partial t}\,{\partial m}} \,+\,
{{2\,{\rm e}^{\phi_\circ}}\over r_0}\,{{\partial r_2}\over{\partial t}} \,-\,
{{2\,r_1\,{\rm e}^{\phi_\circ}}\over r_0^2}\,{{\partial r_1}\over{\partial t}}
\,-\,r_0^4\,{\rm e}^{-\phi_\circ}\,{{\partial r_1}\over {\partial m}}\,
{{\partial^2 r_1}\over {\partial t}\,{\partial m}} \,\,.
\end{equation}
After using equation~(33) for $\phi_1$ and equation~(35c) for 
$\partial\phi_1/\partial t$, we also have
\begin{mathletters}
\begin{eqnarray}
{\rm e}^{\phi_\circ}\,\phi_1\,{{\partial \phi_1}\over{\partial t}} 
&\,=\,& \left(
r_0^2\,{\rm e}^{-\phi_\circ}\,{{\partial r_1}\over{\partial m}} \,+\,
{{2\,r_1}\over r_0}\right)
{{\partial {\phantom u}}\over {\partial m}} 
\!\left( r_0^2\,{{\partial r_1}\over {\partial t}}\right) \,\,, \\
{\phantom u} &\,=\,& 
2\,r_0\,{{\partial r_1}\over {\partial m}}\,{{\partial r_1}\over {\partial t}} 
\,+\, r_0^4\, {\rm e}^{-\phi_\circ}\,{{\partial r_1}\over {\partial m}}\,
{{\partial^2 r_1}\over {\partial t}\,{\partial m}} \,+\,
4\,r_1\,{{\partial r_0}\over {\partial m}}\,{{\partial r_1}\over {\partial t}}
\,+\, 2\,r_0\,r_1\,{{\partial^2 r_1}\over {\partial t}\,{\partial m}} \,\,,
\end{eqnarray}
\end{mathletters}
where we have again utilized equation~(20). Combining equations~(46) and (47b) 
then gives
\begin{equation}
{\rm e}^{\phi_\circ} 
\left( \phi_1\, {{\partial\phi_1}\over {\partial t}} \,+\,
{{\partial\phi_2} \over {\partial t}}\right) \,=\,
{{\partial {\phantom u}} \over  {\partial m}}\! 
\left( 2\,r_0\,r_1\,{{\partial r_1}\over {\partial t}} \,+\,
r_0^2\,{{\partial r_2}\over {\partial t}}\right)
\,\,, 
\end{equation} 
which is equivalent to equation~(24).

Except for the inhomogeneous term $F$, equation~(44) for $r_2 (m,t)$ is 
identical to equation~(34) for $r_1 (m,t)$. It is precisely because of this 
extra term that the second-order motion {\it cannot} be described as a normal 
mode of oscillation. Once we single out the zero-frequency, fundamental mode 
for the first-order motion, $F$ can be written purely as a function of $m$:
\begin{equation}
F \,=\, \left( {2\over r_0^3}\,-\, {{3\,m}\over r_0^4}\right)\xi_1^2 \,+\,
{\rm e}^{-\phi_\circ}\left( {{2\,m}\over r_0}\,-\,2\right)
{{d\xi_1^2}\over{dm}} \,+\,
r_0^2\,{\rm e}^{-2\,\phi_\circ} \left( 2\,r_0\,-\,m\right)
\left( {{d\xi_1}\over{dm}}\right)^2 \,\,.
\end{equation}

The boundary conditions on $r_2 (m,t)$ are the usual ones:
\hbox{$r_2 (0,t)\,=\,0$} and \hbox{$\phi_2 (M,t) \,=\,0$}. From equation~(43),
the outer boundary condition is more usefully recast as
\begin{mathletters}
\begin{eqnarray}
{{\partial r_2}\over {\partial m}}  &\,=\,&
{{{\rm e}^{\phi_\circ}}\over r_0^2}  
\left({{3\,r_1^2}\over r_0^2} \,-\, {{2\,r_2}\over r_0}\right)
\,\,, \\ 
{\phantom u} &\,=\,&  
{{{\rm e}^{\phi_\circ}}\over r_0^2} 
\left({{3\,\xi_1^2}\over r_0^2} \,-\, {{2\,r_2}\over r_0}\right)
\,\,.
\qquad{\it at}\,\,\, m\,=\,M
\end{eqnarray}
\end{mathletters}

\subsection{Method of Solution}

Solving the partial differential equation~(44) requires that we first specify
the full initial state of the cloud. Assuming the fundamental mode has zero
frequency, i.e., that \hbox{$r_1\,=\,\xi_1$}, we still need to set the 
functional form of $r_2 (m,0)$. As our fiducial initial state, we demand that 
the cloud's density profile be just that resulting from the normal mode acting 
on the equilibrium state. That is, we set \hbox{$\phi_2 (m,0) \,=\,0$}. With 
this condition, equation~(17) then gives the physical interpretation of the 
parameter $\epsilon$ as the nondimensional amplitude of the initial 
$\phi$-perturbation.

The vanishing of $\phi_2 (m,0)$ does {\it not} mean that 
\hbox{$r_2 (m,0) \,=\, 0$}. Instead, the initial $r_2$-profile follows by 
setting \hbox{$\phi_2\,=\,0$} in equation~(43) and specializing to
\hbox{$r_1 \,=\, \xi_1$}:
\begin{equation}
r_0^2\,{\rm e}^{-\phi_\circ}\,{{dr_2}\over{dm}}
\,+\,{{2\,r_2}\over r_0} \,=\, {\xi_1^2\over r_0^2} \,+\,
{{r_0^4\,{\rm e}^{-2\,\phi_\circ}}\over 2}\,
\left({{d\,\xi_1}\over{dm}}\right)^2 \,\,.
\qquad{\it at}\,\,\, t\,=\,0 
\end{equation}
We solve this ordinary differential equation for $r_2 (m,0)$. After again 
noting the regular singular point at the origin, we begin the numerical
integration through a power-law expansion:
\begin{equation}
r_2 \,=\, d_0\,m^{1/3} \,+\, d_1\,m \,+\, d_2\,m^{5/3} \,+\, ... \,\,.
\end{equation}
The expansion coefficients are readily found:
\begin{equation}
d_0 \,=\, {1\over{2\,a_0}} \,\,.
\qquad {\it etc.} 
\end{equation}
The top curve of Figure~3 displays the calculated $r_2 (m,0)$, again as a 
function of $r_0$.

We also need to specify the initial velocity 
\hbox{$\partial r_2/\partial t \,(m,0)$}. An interesting, and physically 
relevant situation is when the cloud is perfectly static, i.e., when
\begin{equation}
{{\partial r_2}\over{\partial t}} \,=\, 0 \,\,.
\end{equation}
More realistically, the cloud has slowly evolved from some earlier state,
perhaps through the effect of ambipolar diffusion. In the absence of a more
complete evolutionary picture, we shall adopt the simplest, zero-velocity,
initial condition.

Equation~(44) may be solved numerically through the method of characteristics. 
For this purpose, it is convenient to adopt $r_0$ as the independent, spatial 
variable. Using the connection between $r_0$ and $m$ in equation~(20), the 
fundamental equation~(44) becomes
\begin{equation}
{{\partial^2 r_2}\over{\partial t^2}}\ \,=\, 
{{\partial^2 r_2}\over{\partial r_0^2}} \,+\,
\left( {2\over r_0} \,-\, {m\over r_0^2} \right)
{{\partial r_2}\over {\partial r_0}} \,+\,
\left( {{2\,m}\over r_0^3} \,-\, {2\over r_0^2}\right) r_2 \,+\, F \,\,,
\end{equation}
where $F$ is now written as
\begin{equation}
F \,=\, \left( {2\over r_0^3}\,-\, {{3\,m}\over r_0^4}\right) \xi_1^2 
\,+\, \left( {{2\,m}\over r_0^3}\,-\,{2\over r_0^2}\right) 
{{d\xi_1^2}\over{dr_0}} \,+\,
\left( {2\over r_0}\,-\,{m\over r_0^2}\right)
\left( {{d\xi_1}\over{dr_0}}\right)^2 \,\,.
\end{equation}

The reason for changing Lagrangian variables from $m$ to $r_0$ is that the 
characteristics of equation~(55) are simply
\begin{equation}
{{d r_0}\over{dt}} \,=\, \pm\,1 \,\,.
\end{equation} 
In light of our nondimensionalization, we see that small disturbances in the 
cloud travel at the isothermal sound speed, an intuitively appealing result. We
solve equation~(55) by propagating the partial derivatives
\hbox{$\partial r_2/\partial r_0$} and \hbox{$\partial r_2/\partial t$} along
the $(r_0, t)$ grid, as detailed in the Appendix. The starting value of
\hbox{$\partial r_2/\partial t$} is everywhere zero, according to 
equation~(54). As we advance in time, we also incorporate both the inner
boundary condition \hbox{$r_0\,=\,0$} and the outer one of equation~(50b). The
latter is now more conveniently written as
\begin{equation}
{{\partial r_2} \over {\partial r_0}} \,=\, 
{{3\,\xi_1^2} \over r_0^2} \,-\,
{{2\,r_2}\over r_0} \,\,.
\qquad{\it at}\,\,\, m\,=\,M
\end{equation}

\subsection{Numerical Results}

Integration of the partial differential equation~(55) for $r_2 (r_0,t)$ shows
that this function globally decreases for \hbox{$t > 0$}. That is, all mass
shells contract, although the cloud was neither expanding nor contracting in 
its initial state. If this state were slightly compressed from equilibrium,
the parameter $\epsilon$ would be negative. If slightly expanded, $\epsilon$ 
would be positive. In either case, the second-order effects governing 
subsequent evolution are proportional to $\epsilon^2$. Thus, the cloud
inevitably contracts.

This contraction begins slowly, but accelerates with time. To illustrate the
effect, Figure~3 displays $r_2 (r_0,t)$ for \hbox{$t\,=\,0$}, 1, and 2. We see
how any initial, slight inflation (assuming \hbox{$\epsilon\,>\,0$}) is
everywhere reversed by \hbox{$t\,=\,1$}. By \hbox{$t\,=\,2$}, mass shells 
roughly midway from the center are clearly contracting the most rapidly.

The time span shown in the figure exceeds that associated with free-fall 
collapse. The latter is conventionally taken to be
\begin{equation}
t_{\rm ff} \,=\, \sqrt{{3\,\pi}\over{32\,G\,\rho_c}} \,\,,
\end{equation}
where $\rho_c$ is the cloud's central density. (Here, we are temporarily 
reverting to dimensional variables.) Our nondimensional time $\tilde t$ is
given by equation~(12), so that
\begin{equation}
{t\over{t_{\rm ff}}} \,=\, {1\over\pi}
\sqrt{{8\,\rho_c}\over{3\,\rho_e}}\,\,{\tilde t} \,\,.
\end{equation}
Here, \hbox{$\rho_e\,\equiv\,P_e/a^2$}. For the Bonnor-Ebert initial state 
(\hbox{$\rho_c/\rho_e\,=\,14.04$}), the numerical coefficient in equation~(60) 
is 1.95. Thus, the last profile shown in Figure~3 corresponds to 
\hbox{$2\times 1.95\,=\,3.90$} free-fall times.

We are primarily interested, not in the displacement of mass shells, 
but in the contraction speed. From equation~(18), the nondimensional velocity 
(i.e., its value relative to the sound speed $a$) is given by 
\hbox{$\epsilon^2\,\partial r_2/\partial t$}. The actual velocity of each mass
shell thus depends on the value of $\epsilon$, which we do not know 
a priori. \citet{k06} have recently matched molecular-line profiles from 
the starless core B68 by assuming the object is undergoing a nonradial 
oscillation mode, with a dimensionless amplitude of 0.25. Using this figure as 
rough guide, we provisionally choose \hbox{$\epsilon\,=\,+0.2$} and display 
the resulting velocity profiles in Figure~4. Here, the times are identical to 
those in Figure~3. However, the independent spatial variable is now the full 
radius $r$, as given by equation~(18). 

We see that the velocity of all mass shells is negative, with a magnitude that
increases with time. Thus, contraction is indeed accelerating. By the last time
shown, the largest contraction speed within the cloud is about 0.2 times the 
sound speed. Interestingly, the cloud edge also has negative velocity, 
suggesting that contraction is spreading into the exterior region. 

Closer to the cloud center, it appears from Figure~4 that the velocity at any 
time increases linearly with radius. This feature of the evolution was found by
\citet{fc93} in their numerical simulation. We may derive the result 
analytically by examining the fundamental equation~(55) in this limit. Consider
first the inhomogeneous term $F$. Now the series expansion for $\xi_1 (m)$ in 
equation~(39) tells us that $\xi_1\,=\,e_0\,r_0$ to lowest order, where $e_0$
is a constant. From equation~(56), $F$ has three terms that diverge as $r_0$ 
approaches zero. Their sum is
\begin{equation}
{{2\,\xi_1^2 \over r_0^3}} \,-\,
{2\over r_0^2} \,{{d\xi_1^2}\over dr_0} \,+\,
{2\over r_0} \left({{d\xi_1}\over{dr_0}}\right)^2 \,=\,
{{2\,e_0^2}\over r_0} \,-\,
{{4\,e_0^2}\over r_0} \,+\,
{{2\,e_0^2}\over r_0} \,\,,
\end{equation} 
which vanishes. However, the rest of equation~(55) has coefficients that still
do diverge. Close to the origin, the equation reduces to
\begin{equation}
0 \,=\, {{\partial^2 r_2}\over{\partial r_0^2}} \,+\,
{2\over r_0}\,{{\partial r_2}\over{\partial r_0}} \,-\,  
{{2\,r_2}\over r_0^2}\,\,.
\qquad {r_0 \rightarrow 0} 
\end{equation}
Here we have used the fact that, since \hbox{$r_2 \,=\, 0$} at the origin for 
all times, we may neglect the second time derivative on the left side of
equation~(55). Since equation~(62) itself holds for all times, we may take
its time derivative and obtain the analogous relation for the velocity
\hbox{$v\,\equiv\,\epsilon^2\,\partial r_2/\partial t$}:
\begin{equation}
0 \,=\, {{\partial^2 v}\over{\partial r_0^2}} \,+\,
{2\over r_0}\,{{\partial v}\over{\partial r_0}} \,-\,  
{{2\,v}\over r_0^2}\,\,.
\qquad {r_0 \rightarrow 0} 
\end{equation}
The non-divergent solution to this equation is that $v$ is indeed proportional 
to $r_0$.

Figure~5 shows the evolution of the cloud's density profile $\rho (r,t)$ over
the same time interval as in Figures~3 and 4. Since $\epsilon$ was assumed to 
be positive, the cloud begins in a slightly inflated state, with a central 
density of only 5.15. However, subsequent contraction drives up the density, 
which reaches a central value of 15.5 by \hbox{$t\,=\,2$}. Since all internal 
velocities are still subsonic at this time, the density profile is consistent 
with a cloud that, to first order, is in hydrostatic balance. Force balance 
will, of course, be strongly violated in the future, as the cloud enters a 
state of true collapse.

The perturbative nature of our calculation limits us to describing the initial 
transition phase. Our calculation is only valid to the point where second-order
terms in the expansions become comparable to their first-order counterparts.
For \hbox{$\epsilon\,=\,+0.2$}, the maximum absolute value of
$\epsilon^2\,r_2 (m,t)$ in equation~(18) becomes equal to the maximum value
of $\epsilon\,r_1 (m,t)$ at \hbox{$t\,=\,1.9$}. Thus, the calculation is still 
marginally valid for the \hbox{$t\,=\,2$} profiles shown in Figures~3, 4, and 
5, but not beyond those.  

\subsection{Alternative Initial Condition}

It is important to verify that the evolutionary results shown thus far are not
sensitive to the detailed initial state chosen. Recall that our fiducial state
represented a pure first-order perturbation of the density, in that we set
\hbox{$\phi_2\,=\,0$} for all mass shells. This choice was convenient, but
arbitrary. We could have chosen any initial density profile consistent with the
boundary condition \hbox{$\phi_2 (M,t)\,=\,0$}. 

Another starting configuration is obtained by setting
\begin{equation}
\phi_2 (m,0) \,=\, {\rm cos}\,\left({{\pi\,m}\over{2\,M}}\right) \,\,.
\end{equation}
With this $\phi_2$-profile, the outer boundary condition is satisfied, but the
central $\phi_2$ is now unity. Since \hbox{$\rho\,=\,{\rm exp}\,(-\phi)$}, the
initial cloud is less dense than before. If we insert equation~(61) into
equation~(43), we may again integrate an ordinary differential equation for
$r_2 (m,0)$. As shown in the left panel of Figure~6, the $r_2$-profile is thus
inflated relative to the previous initial state, with a maximum fractional
difference of 43~percent. Nevertheless, subsequent contraction brings the cloud
to a very similar configuration. The right panel of Figure~6 displays the 
new $r_2$-profile at \hbox{$t\,=\,2$}. It differs from the one obtained using
the original starting state by at most 8~percent. The profile of velocity,
\hbox{$\epsilon^2\,\partial r_2/\partial t$}, is also very close to the
previous one, with the maximum speeds differing by less than 1~percent at 
\hbox{$t\,=\,2$}. 

\section{Discussion}

In the picture introduced here, star-forming dense cores undergo a prolonged
phase of contraction before their ultimate collapse. The contraction is slow
because the cloud's self-gravity is still nearly counteracted by the outward
pressure gradient. The slight imbalance of these forces creates subsonic,
inward motion that gradually accelerates. Previous researchers performing
direct simulations have also documented slow motion prior to runaway collapse.
However, the characteristics of this phase, and indeed whether it led to
collapse or rebound, depended on the unavoidable artifices of a direct 
simulation: the imposed overdensity in the cloud, the numerical accuracy of 
the initial state, and the precise treatment of the central few zones 
\citep{h77,bb82,fc93,o99}. We have demonstrated physically how an initially 
static, marginally stable cloud inevitably contracts. This contraction is 
essentially the inward phase of a slow oscillation that smoothly leads to 
free-fall collapse.

Motivated by the observed infall signature of many starless dense cores, 
others have offered different pictures. \citet{kf05} hypothesized that 
some cores are born in a gravitationally unstable state. They followed the
collapse of such an object numerically, accounting for internal cooling by
molecular lines and thermal dust emission. This work extends that of
\citet{z90}, who modeled the (starred) core B335 as undergoing collapse from a
singular isothermal sphere.\footnote{The singular isothermal sphere is 
unstable not only to the fundamental oscillation mode, but to all higher
harmonics \citep[][Chapter 9]{sp04}. The collapse of an unbounded, singular
sphere was calculated in a self-similar fashion by \citet{s77}. In light of
the infall observations, \citet{f04} generalized this model to include a 
finite, inward velocity at large radii.} As already noted, the early history 
of unstable objects is problematic; \citet{kf05} speculate that the cloud 
fragmented from a larger, turbulent velocity field. \citet{ml98} attributed
this localized fragmentation to enhanced dissipation via ion-neutral friction,
leading to a pressure-driven, inward flow.

Our general concern about using unstable or actively collapsing states to
match observations is their brevity. To illustrate the point more 
quantitatively, suppose the infall signatures reflect {\it first}-order motion,
i.e., that the cloud's fundamental eigenfrequency is nonzero. The 
nondimensional perturbation $r_1 (m,t)$ would then be expressed as
\begin{equation}
r_1 \,=\, -\xi_1\,{\rm e}^{+|\omega_\circ|t} \,\,.
\end{equation}
The function $\xi_1 (m)$ obeys equation~(37), but with $\omega^2$ replaced by
$-|\omega_\circ|^2$. If we again normalize the coefficient $c_\circ$ in
equation~(39) to unity, then $\xi_1 (m)$ resembles the curve $\xi_{10}$ in
Figure~2. Notice the overall minus sign in equation~(65) that ensures 
contraction prior to free-fall collapse. Over what time does this transition 
occur?

The physical velocity, in units of the sound speed, is
\hbox{$\epsilon\,\partial r_1/\partial t \,=\, \epsilon\,|\omega_\circ|\,r_1$}.
If $\epsilon$ is still about 0.2, and if we are to match velocities of 0.2 
times the sound speed at the present epoch (\hbox{$t\,=\,0$}), then the growth 
rate $|\omega_\circ|^{-1}$ must be about unity. The {\it dimensional} time $t$ 
for the velocity to increase by a factor $e$ is, from equations~(9) and (12),
\begin{equation}
t \,=\,{{G\,M}\over{{\tilde M}\,a_t^3}} \,\,.
\end{equation} 
The numerical solution to the altered equation~(37) tells that the 
nondimensional cloud mass $\tilde M$ corresponding to 
\hbox{$|\omega_\circ|\,=\,1$} is 4.02. The center-to-edge density contrast of 
this object is 30, a figure that is marginally consistent with observations
(recall Fig.~1). Returning to L1544, its gas temperature is 10~K \citep{bm89},
while the most recent mass estimate is $2\,\,\Msun$ \citep{oh99}. Equation~(66)
then gives an $\rm e$-folding time of $2\times 10^5$~yr. The statistical 
prevalence of starless dense cores makes it unlikely that they are evolving 
over such a brief, dynamical interval. Note finally that if the initial 
perturbation amplitude $\epsilon$ were smaller than 0.2, then $|\omega_\circ|$ 
would be correspondingly larger, driving down the evolutionary time even more.

We emphasize again the general nature of our theoretical finding. A more 
realistic model for a starless core should certainly include the anisotropic 
supporting force from the interstellar magnetic field. Such magnetostatic 
configurations are themselves subject to global oscillations. The lowest 
eigenfrequency vanishes in the marginally stable state \citep{t88}. These 
states, analogues of the Bonnor-Ebert configuration used here, will also 
undergo slow contraction prior to collapse, even under the approximation of 
flux freezing. They are close to being magnetically supercritical, so that 
ambipolar diffusion may enhance the contraction process \citep{cb01}. In any 
event, it will be interesting to track the evolution through the transonic 
phase into full collapse, both in our spherical model and its magnetized 
generalization. 

Returning to the observational motivation of this study, it will also be
instructive to calculate, within our spherical model, the predicted profiles 
for molecular emission lines of varying optical depth. For our representative
case of \hbox{$\epsilon\,=\,+0.2$}, we find a maximum infall speed of 0.2 times
the sound speed after about 4 free-fall times. This speed is comparable to 
the 0.08~km~s$^{-1}$ inferred for the well-studied starless core L1544 through
N$_2$H$^+$ observations \citep{wi99} We are encouraged by this finding, but 
stress the need for more comprehensive and detailed comparisons. Note 
especially that the amplitude $\epsilon$ is not readily observable with any 
precision; nor is the evolutionary time $t$. And yet contraction models of 
lower $\epsilon$ and larger $t$ broadly mimic, in their velocity profiles, 
those with higher $\epsilon$ and shorter $t$. Hopefully, calculated line 
profiles will differ enough to resolve this ambiguity. If some of the profiles
successfully match observations, we will not only have gained new insight into
the mechanism of dense core contraction, but also a new measure for their 
longevity prior to collapse.

\acknowledgments

We are grateful for stimulating conversations with Phil Chang and Steve Shore
during the inception of this project. We also thank the referee, Tom Hartquist,
for comments that improved the original manuscript. S. S. was partially 
supported by NSF Grant AST-0639743.

\clearpage

\appendix

\section{Implementing the Method of Characteristics}
Equation~(55) is a linear, inhomogeneous partial differential equation in the
independent variables $r_0$ and $t$. Following standard procedure 
\citep[e.g.,][]{a92}, we suppose that along some curve in the \hbox{$(r_0, t)$}
plane, we know $r_2$ and its first partial derivatives 
\hbox{$\partial r_2/\partial r_0$} and \hbox{$\partial r_2/\partial t$}. Then
the three second partial derivatives are related through equation~(55), which
we rewrite as
\begin{equation}
{{\partial^2 r_2}\over{\partial t^2}} \,-\,
{{\partial^2 r_2}\over{\partial r_0^2}} \,=\, g \,\,.
\end{equation}
Here,
\begin{equation}
g\,\equiv\,
\left({2\over r_0}\,-\,{m\over r_0^2}\right) 
{{\partial r_2}\over{\partial r_0}} \,+\,
\left({{2\,m} \over r_0^3} \,-\, {2\over r_0^2}\right) r_2 \,+\, F \,\,,
\end{equation}
and $F$ is given by equation~(56). Along our curve, we also have the
differential relations
\begin{eqnarray}
{{\partial^2 r_2}\over{{\partial r_0}\,{\partial t}}} \,\,\Delta t \,+\,
{{\partial^2 r_2}\over{\partial r_0^2}} \,\,\Delta r_0 &\,=\,&
\Delta \!\left({{\partial r_2}\over{\partial r_0}}\right) \,\,, \\
{{\partial^2 r_2}\over{\partial t^2}} \,\,\Delta t \,+\,
{{\partial^2 r_2}\over{{\partial r_0}\,{\partial t}}} \,\,\Delta r_0 &\,=\,&
\Delta \!\left({{\partial r_2}\over{\partial t}}\right) \,\,.
\end{eqnarray}

Equations~(A1), (A3), and (A4) constitute three algebraic relations for the
second partial derivatives. We may recast the system in matrix form
\begin{equation}
M\,{\bx} \,=\, {\by} \,\,,
\end{equation}
where
\begin{equation}
\bx \,=\, \left(
\begin{array}{c}
\partial^2 r_2/\partial t^2  \\
\partial^2 r_2/\partial r_0 \,\partial t \\
\partial^2 r_2/\partial r_0^2   
\end{array}
\right) \,\,,
\end{equation}

\begin{equation}
\by \,=\, \left(
\begin{array}{c}
g  \\
\Delta ({\partial r_2/\partial r_0})   \\
\Delta ({\partial r_2/\partial t}) \,\,
\end{array}
\right) \,\,,
\end{equation}
and
\begin{equation}
M \,=\, 
\left(
\begin{array}{ccc}
1 & 0 & -1 \\
0 & \Delta t & \Delta r_0  \\
\Delta t & \Delta r_0 & 0
\end{array}
\right) \,\,.
\end{equation}
The vector $\bx$ is uniquely determined {\it unless} 
\hbox{${\rm det}\,M\,=\,0$}. Thus, discontinuities propagate along 
characteristics given by
\begin{equation}
{{\Delta r_0}\over{\Delta t}} \,=\, \pm 1 \,\,,
\end{equation} 
as in equation~(57).

We may also replace any column in $M$ by the vector $\by$. For example, 
consider the matrix $M^\prime$ given by
\begin{equation}
M^\prime \,=\, 
\left(
\begin{array}{ccc}
1 & g & -1 \\
0 &  \Delta (\partial r_2/\partial r_0) & \Delta r_0 \\
\Delta t  &  \Delta (\partial r_2/\partial t) & 0
\end{array}
\right) \,\,.
\end{equation}
There is no solution at all for the second derivatives unless
\hbox{${\rm det}\,M^\prime \,=\,0$}. This condition gives us the further
differential relation
\begin{equation}
\Delta \!\left({{\partial r_2}\over{\partial r_0}}\right) \,-\, 
\Delta \!\left({{\partial r_2}\over{\partial t}}\right) \,=\,
g\,\Delta r_0 \,\,,
\end{equation}
along the +~characteristic, and 
\begin{equation}
-\Delta \!\left({{\partial r_2}\over{\partial r_0}}\right) \,-\, 
\Delta \!\left({{\partial r_2}\over{\partial t}}\right) \,=\,
g\,\Delta r_0 \,\,,
\end{equation}
along the -~characteristic.

Figure~7 is a schematic spacetime diagram that illustrates the practical
procedure. We set up uniform grids along the $r_0$- and $t$-axes. At 
\hbox{$t\,=\,0$}, we have \hbox{$\partial r_2/\partial t \,=\,0$}, while the
solution to equation~(51) gives the initial values of 
\hbox{$\partial r_2/\partial r_0$} for the fiducial initial state.
(An analogous equation is solved for the alternative state; see \S  4.4.) 
Starting from any two adjacent points on the $r_0$-axis, such as $A$ and $B$, 
we use equations~(A11) and (A12) to solve simultaneously for the two first 
partial derivatives at point $D$, where the two characteristics intersect.
Similarly, we find the two derivatives at $E$ starting from the pair $B$ and
$C$. The derivatives at $D$ and $E$ then yield those at point $G$, and so on.  

We also need to propagate the information contained in the central and surface
boundary conditions. Since \hbox{$r_2 (0,t) \,=\,0$}, it is also true that
\hbox{$\partial r_2/\partial t\,=\, 0$} at points such as $F$ in the figure. 
This condition, along with equation~(A12) for the -~characteristic
joined to point $D$, allows us to solve for \hbox{$\partial r_2/\partial r_0$}
at $F$. Since both partial derivatives are now known at $F$ and $G$, this
information can then be used to find the derivatives at points further advanced
in time. Near the cloud's outer edge, we establish the two partial derivatives
at point $L$ in the usual manner. Knowing this information at $L$, we find one
relation between the two derivatives at the boundary point $N$ by using 
equation~(A11) for the +~characteristic. The outer boundary condition,
equation~(58) then supplies a second relation between the derivatives. To find
$r_2$ itself at $N$, we further use
\begin{equation}
\Delta r_2 \,=\, {{\partial r_2}\over{\partial t}}\,\Delta t \,+\,
{{\partial r_2}\over{\partial r_0}}\,\Delta r_0 \,\,. 
\end{equation}

\clearpage

\begin{figure}
\plotone{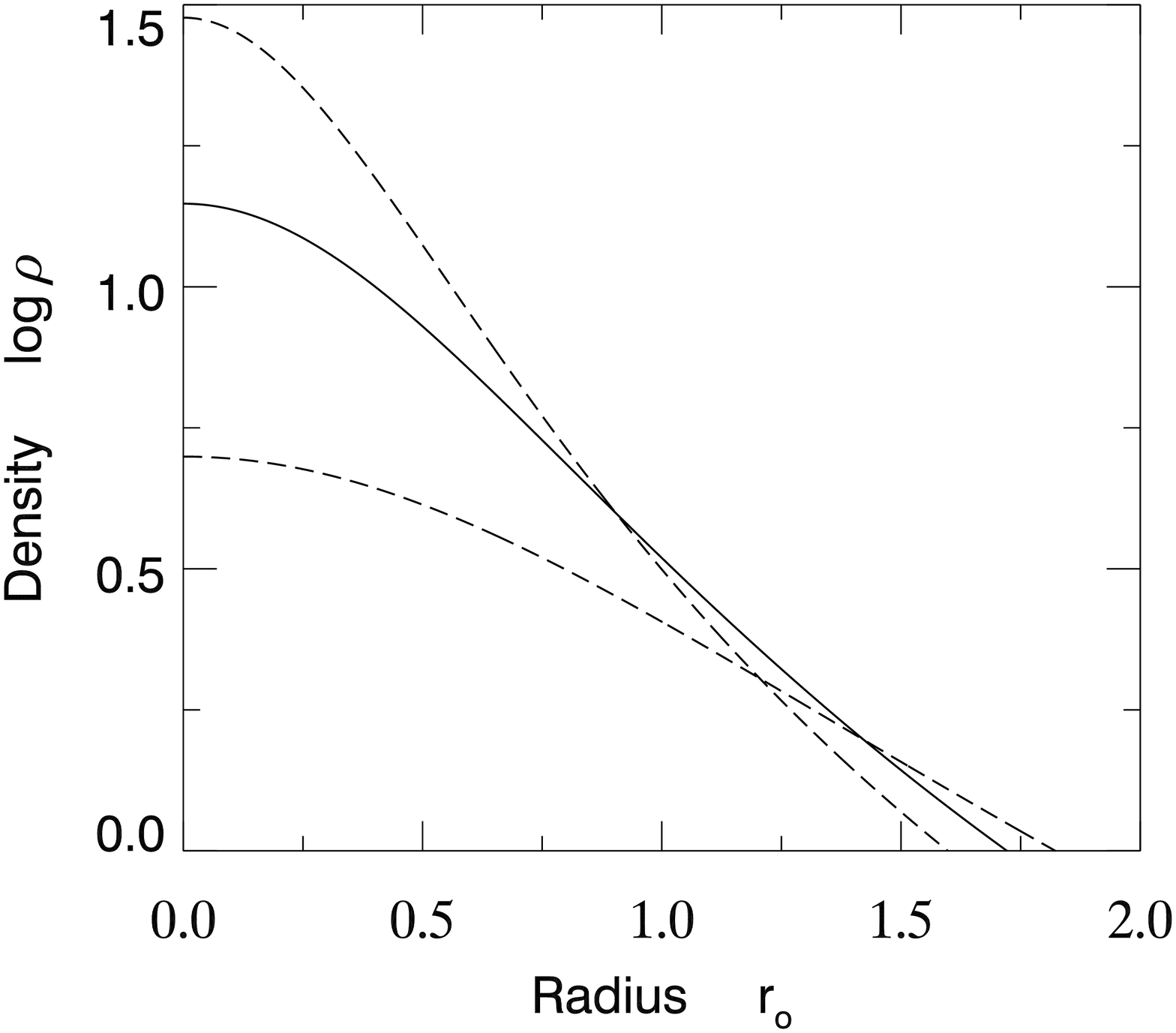}
\caption{Density profiles of equilibrium, isothermal clouds. Both the density
and radius are in the nondimensional units defined in the text. The solid 
curve is the critical Bonnor-Ebert state, while the two dashed curves represent
approximate outer bounds obtained by empirical fitting to starless dense
cores.}
\end{figure}

\begin{figure}
\plotone{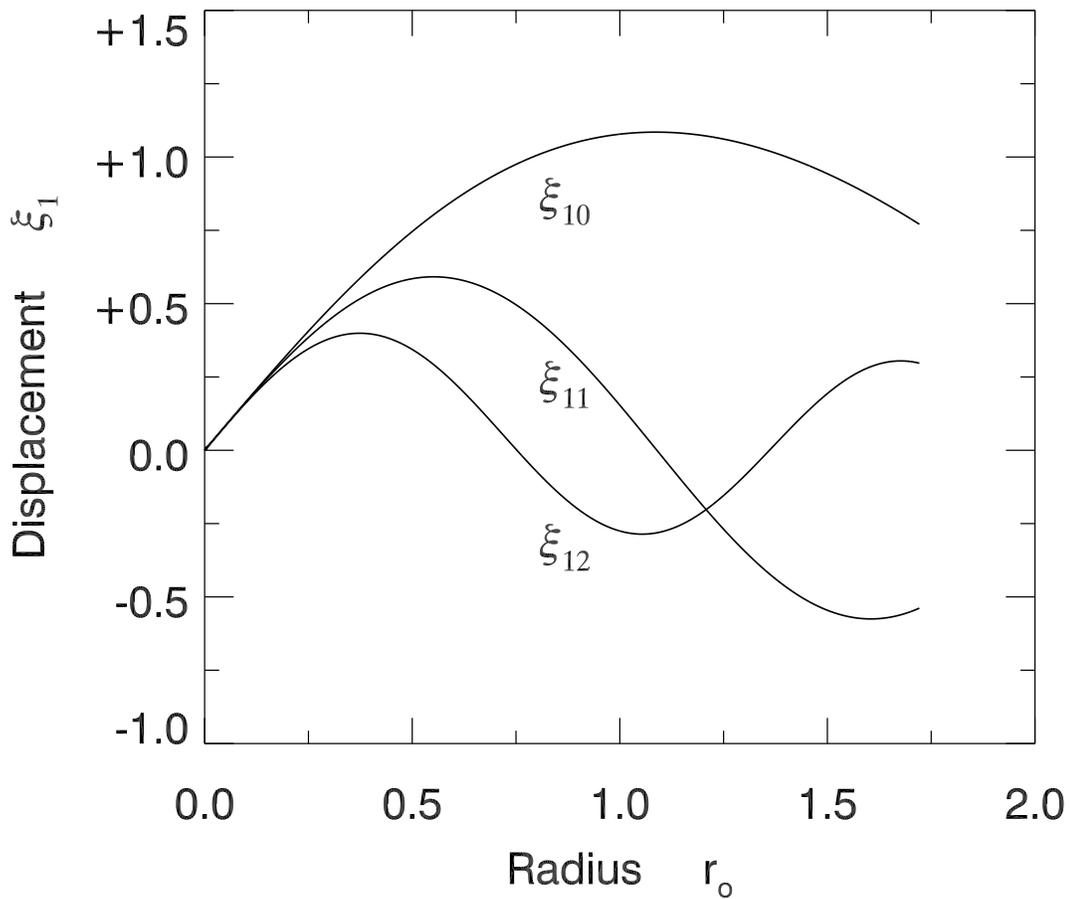}
\caption
{Normal modes of the Bonnor-Ebert isothermal sphere. Shown is the first-order
displacement $\xi_1$ as a function of radius, for the primary ($\xi_{10}$),
first harmonic ($\xi_{11}$), and second harmonic ($\xi_{12}$). All curves have 
been normalized to have the same initial slope.}
\end{figure}

\begin{figure}
\plotone{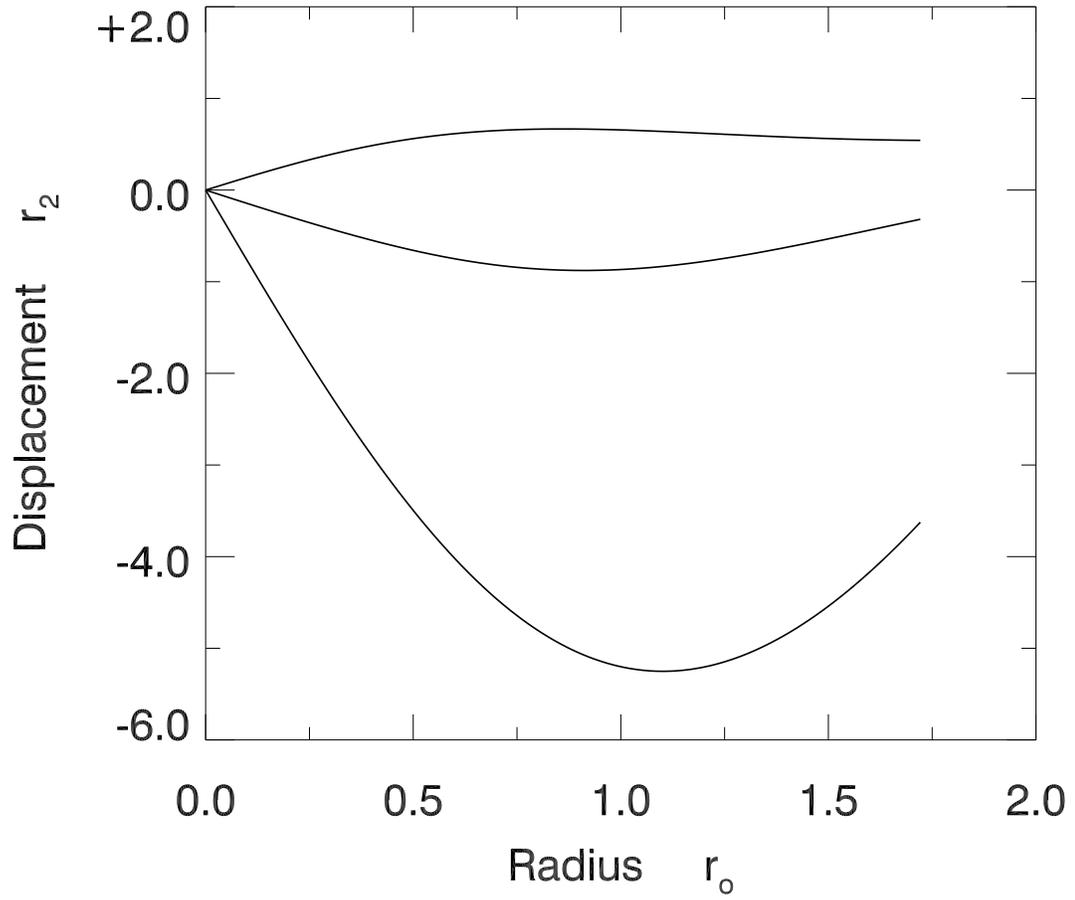}
\caption
{Second-order displacement $r_2$, shown as a function of the Lagrangian radius
$r_0$. From top to bottom, the three profiles correspond to $t\,=\,0$, 1, and
2, respectively}
\end{figure}

\begin{figure}
\plotone{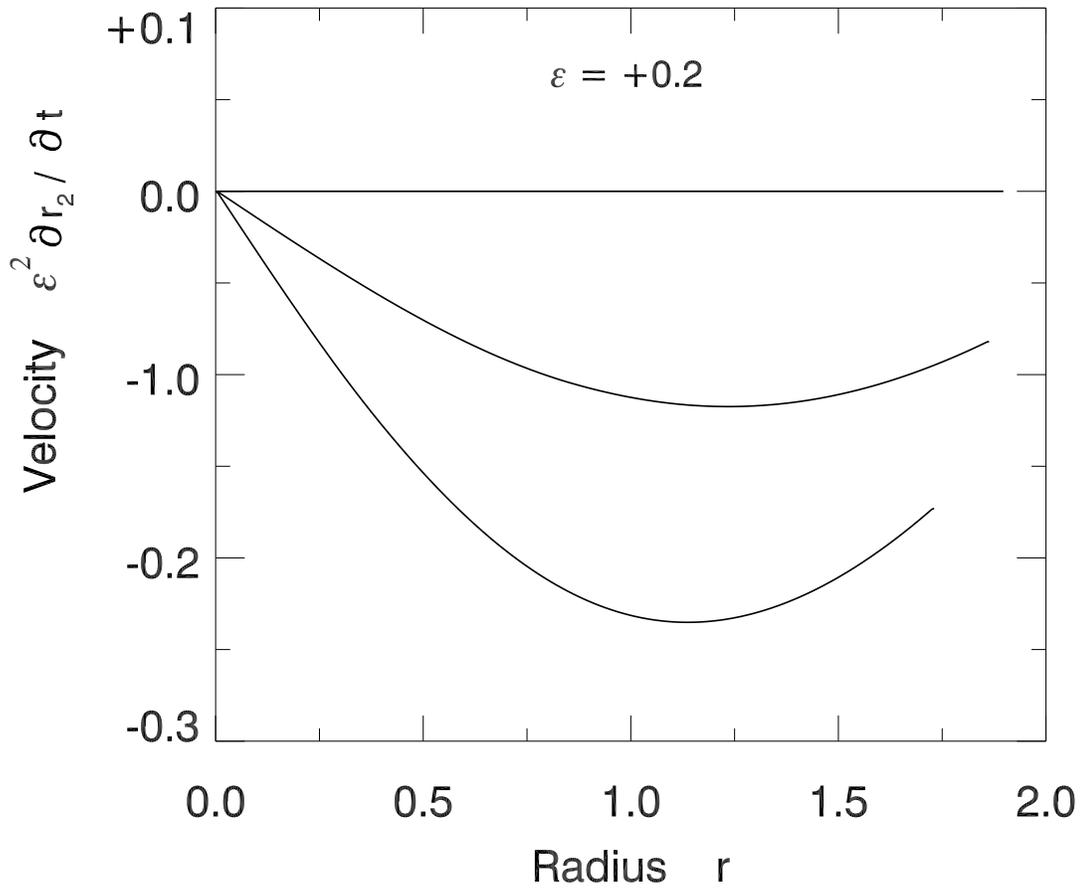}
\caption
{Nondimensional velocity profiles as a function of the full radius $r$. From
top to bottom, the corresponding times are $t\,=\,0$, 1, and 2. The initial
state was perturbed with a dimensionless amplitude \hbox{$\epsilon\,=\,+0.2$}.}
\end{figure}

\begin{figure}
\plotone{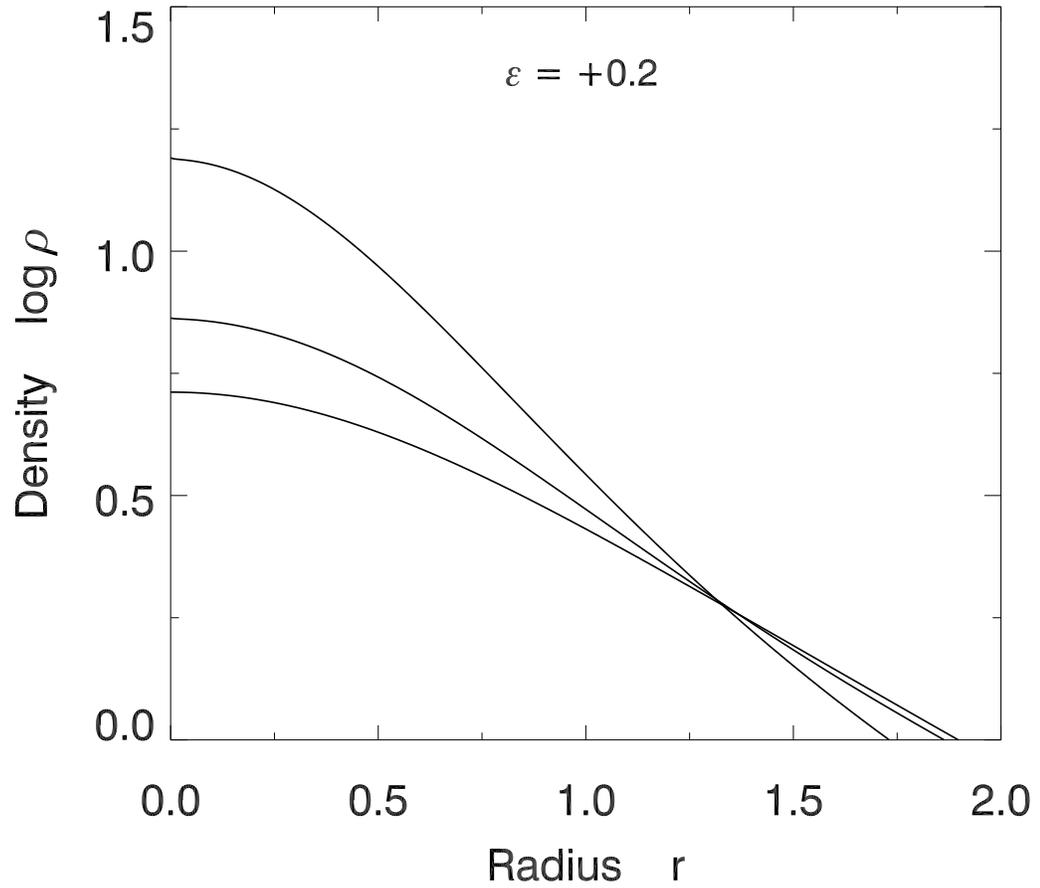}
\caption
{The evolving density profiles for a cloud perturbed initially with
\hbox{$\epsilon\,=\,+0.2$}. The central density monotonically increases for
the three times shown: \hbox{$t\,=\,0$}, 1 and 2.}
\end{figure}

\begin{figure}
\plotone{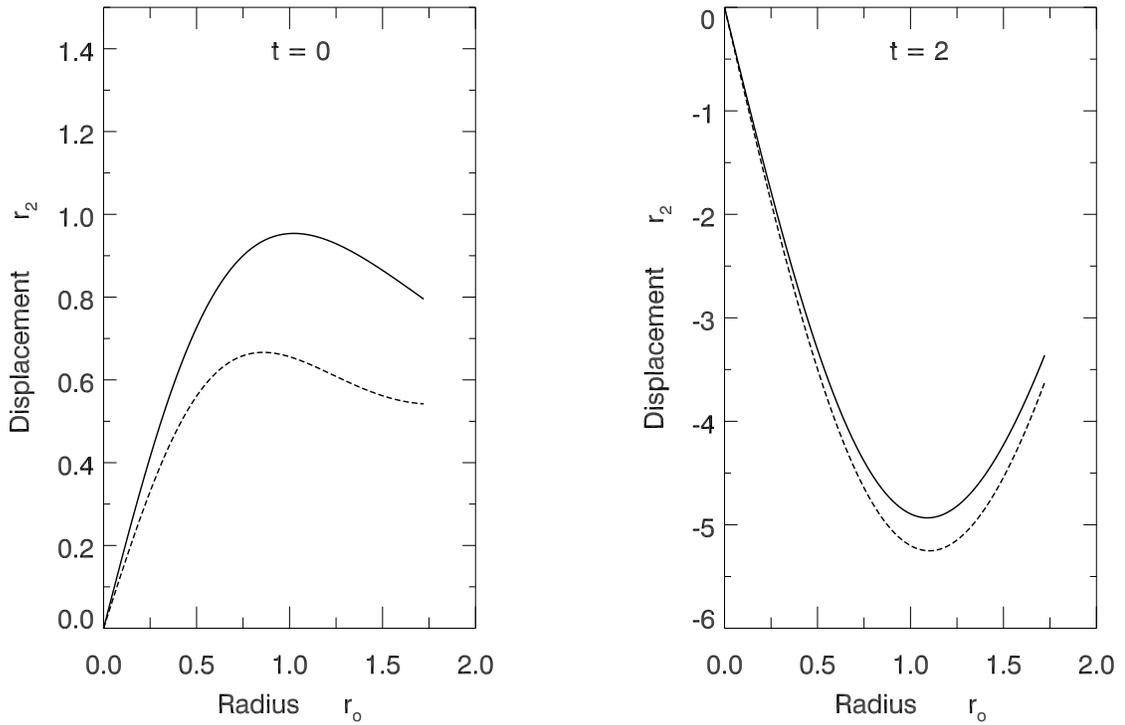}
\caption
{Evolution of the displacement $r_2$ with two initial conditions. The left
panel shows that, at \hbox{$t\,=\,0$}, the $r_2$-profile for the altered 
initial state ({\it solid curve}) differs substantially from that associated 
with the fiducial initial cloud that had a purely first-order density 
perturbation ({\it dashed curve}). Nevertheless, as the right panel shows, the 
profile is close to the previous one at \hbox{$t\,=\,2$}.}
\end{figure}

\begin{figure}
\plotone{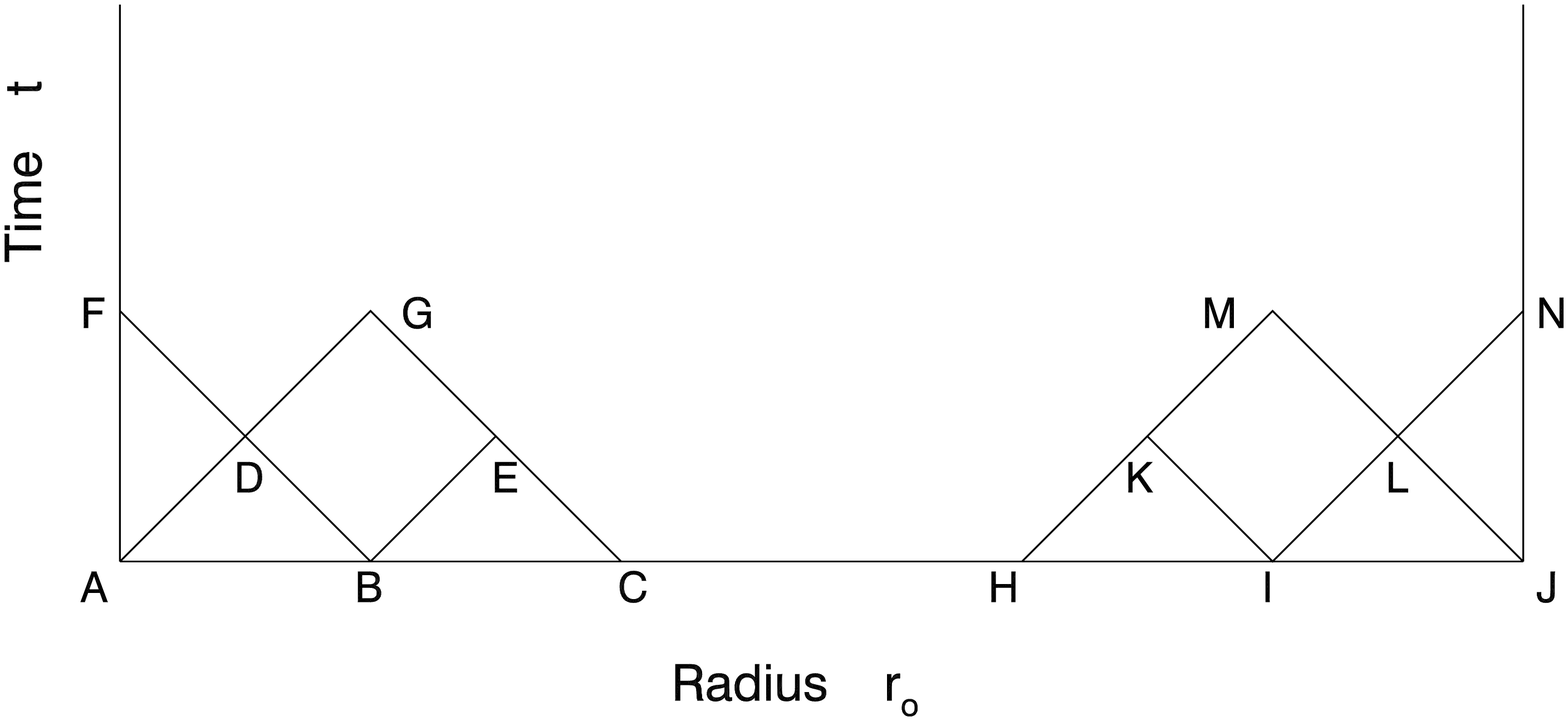}
\caption
{Schematic spacetime diagram illustrating the solution procedure for the
fundamental partial differential equation~(55). The straight, diagonal line 
segment from $A$ to $D$ lies along a +~characteristic, while the line 
segment from $B$ to $D$ lies along a -~characteristic.}  
\end{figure}

\end{document}